\documentclass[preprint,aps,prd,showpacs,nofootinbib,superscriptaddress,tightenlines]{revtex4}

\usepackage{amsfonts}
\usepackage{mathrsfs}
\usepackage{graphicx}
\usepackage{amsmath}
\usepackage{amssymb}
\usepackage{bm}
\usepackage{bbm}
\usepackage{multirow}
\usepackage{alltt}
\usepackage{fancyvrb}
\usepackage[usenames]{color}

\def\PA#1{\left( #1 \right)}
\def\PB#1{\left[ #1 \right]}

\def\N{\mathcal{N}}
\def\Q{\mathcal{Q}}
\def\I{\mathcal{I}}
\def\P{\texttt{Power}}
\def\UV{\rm UV}
\def\nn{\nonumber}

\def\redbox#1{{\color{black}\fbox{$\displaystyle\color{black} #1 $}}}

\begin{document}

\title{A Recursive Method to Calculate UV-divergent Parts at One-Loop Level in Dimensional Regularization}

\author{Feng Feng\footnote{E-mail: fengf@ihep.ac.cn} }
\affiliation{Center for High Energy Physics, Peking University, Beijing 100871, China\vspace{0.2cm}}

\date{\today}

\begin{abstract}
A method is introduced to calculate the UV-divergent parts at one-loop level in dimensional regularization.
The method is based on the recursion, and the basic integrals are just the scaleless integrals after the recursive reduction, which involve no other momentum scales except the loop momentum itself. The method can be easily implemented in any symbolic computer language, and an implementation in \textsc{Mathematica}
is ready to use.
\end{abstract}


\maketitle

{\bf PROGRAM SUMMARY}
\begin{itemize}

\item[]{\em Title of program:} {\tt \$UVPart}
\item[]{\em Programming language:} {\tt \textsc{Mathematica}}
\item[]{\em Available from:}  {\tt \verb=http://power.itp.ac.cn/~fengfeng/uvpart/=}
\item[]{\em Computer(s) for which the program has been designed:}  Any computer where the \textsc{Mathematica} is running.
\item[]{\em External routines/libraries used:} \textsc{FeynCalc}, \textsc{FeynArts}
\item[]{\em Keywords:} UV-Divergences, One-Loop Corrections, Dimensional Regularization
\item[]{\em CPC Library Classification:} {\tt 11.1}
\item[]{\em Nature of problem:} To get the UV-divergent part of any one-loop expression.
\item[]{\em Method of solution:} {\tt \$UVPart} is an \textsc{Mathematica} package where the recursive method has been implemented.
\item[]{\em Running time:} In general it is below a second.

\end{itemize}

\newpage
{\bf LONG WRITE-UP}

\section{Introduction}
One has to deal with an integration over the loop momentum at next-to-leading order, which results to ultraviolet (UV) and infrared (IR) divergencies.
Dimensional regularization\cite{'tHooft:1972fi,'tHooft:1973us} is needed in order to produce meaningful results.
The general one-loop amplitude can be written as of a linear combination of known scalar integrals\cite{'tHooft:1978xw} --- boxes, triangles, bubbles and tadpoles --- multiplied by coefficients that are rational functions of the external momenta and polarization
vectors, plus a remainder which is also a rational function of the latter.

There are many automatic tools available to achieve the general one-loop amplitude, like \textsc{FeynCalc}\cite{Mertig:1990an} and \textsc{FormCalc}\cite{hep-ph/9807565}, which are based on  the traditional Passarino-Veltman\cite{Passarino:1978jh,Denner:1991kt,Denner:2002ii,Denner:2005nn} reduction of Feynman graphs, which can be generated
automatically(FeynArts\cite{Print-90-0144 (WURZBURG),hep-ph/0012260} or QGRAF\cite{Nogueira:1991ex}). In order to produce numerical results, tensor
coefficients functions are calculated using \textsc{LoopTools}\cite{hep-ph/9807565}. For a detailed review, please see Refs.~\cite{hep-ph/9602280,arXiv:0903.4665}.

In the last few years, several groups have been working on the problem
of constructing efficient and automatized methods for the computation of
one-loop corrections for multi-particle processes. Many different interesting
techniques have been proposed: these contain numerical and semi-numerical
methods\cite{hep-ph/0402152,hep-ph/0508308,arXiv:0704.1835,arXiv:0708.2398}, as well as analytic approaches\cite{hep-ph/9409265,hep-ph/9403226,hep-th/0403047,hep-th/0406177} that make use of unitarity
cuts to build NLO amplitudes by gluing on-shell tree amplitudes\cite{hep-ph/0602178,hep-th/0611091}.
For a recent review of existing methods, see Refs.~\cite{arXiv:0707.3342,arXiv:0704.2798}.

Generally, it will be much easier to calculate the UV-divergent parts of the one-loop amplitude than the one-loop amplitude itself, and there are little work on this specific area since one usually need not to calculate this part separately. However if what we are concerned is the renormalization, we need to calculate the UV-divergent parts only, and moreover there are also some cases in which we have to calculate the UV-divergence separately, for example, most packages like \textsc{Fire}\cite{arXiv:0807.3243} and \textsc{Reduze}\cite{Studerus:2009ye}, which implement the integration by parts (IBP) relations\cite{Chetyrkin:1981qh}, treat the scaleless integrals as zero, i.e.
\begin{equation}\label{scaleint1}
\int d^4 k \; (k^2)^n = 0
\end{equation}
however it is well known for the logarithmically divergent scaleless integrals  that
\begin{equation}
\frac{\PA{2\pi\mu}^{4-D}}{i\pi^2} \int d^4 k \; \frac{1}{(k^2)^2} =  \frac{1}{\varepsilon_{\rm UV}} - \frac{1}{\varepsilon_{\rm IR}} \;,
\end{equation}
so Eq.~(\ref{scaleint1}) does not distinguish the UV- and IR-divergence. This will be fine if we consider the amplitude as a whole, since the UV-divergent parts will be canceled by the counter-terms, and the left divergence will only be IR-divergent. But there is no way to know the IR-divergence from a specific Feynman diagram, which is very important when one considers the factorization, where one tries to identity the source of IR- divergence and factorize them out.

A method that allows the extraction of the UV-divergent part of an arbitrary 1-loop tensor N-point coefficient was presented in Ref.~\cite{Sulyok:2006xp}. We want to introduce another simple method to calculate the UV-divergent part in the dimensional regularization.
The method is based on the recursion, and basic integrals are just the scaleless integrals after the recursive reduction, which involve no other momentum scales except the loop momentum itself.
Since the computation in this method just involves algebraic rational operations, so it can be easily implemented in any symbolic computer language.

The paper is organized as follows: We introduce the definitions and notations in Sec.~\ref{secDN}, then describe the calculations in Sec.~\ref{secDC},
and in Sec.~\ref{secMath} we give an implementation with \textsc{Mathematica}, and use this method in a specific process $e^+e^- \to J/\psi + \eta_c$, and finally comes the summary.

\section{Definitions and Notations\label{secDN}}
The general expression associated with UV-divergent parts at one-loop reads
\begin{equation}\label{Qeq}
\mathcal{I} = \frac{\PA{2\pi\mu}^{4-D}}{i\pi^2} \int d^D k \frac{\Q}{\N_1^{n_1} \N_2^{n_2} \cdots \N_{N}^{n_N}}
\end{equation}
with the denominators
\begin{equation}
\N_i = \PA{k+q_i}^2 - m_i^2 + i\epsilon ,
\end{equation}
where $i\epsilon$ denotes the infinitesimally small positive imaginary part, $\mu$ is the reorganization scale, $D$ is the non-integer dimension of space-time defined as $D = 4-2\varepsilon$, $q_i$ are linear combination of external momenta $p_i$, and the numerator $\Q$ is the polynomial of $k^2$ and $k\cdot p_i$.

The following identity about the scaleless integrals is well known in the calculations with dimensional regularization:
\begin{equation}\label{Identity1}
\frac{\PA{2\pi\mu}^{4-D}}{i\pi^2} \int d^D k \PA{ k^2 }^n  =
\left\{ \begin{array}{ll}
0 & ( n \ne -2 ) \\
\frac{1}{\varepsilon_{\rm UV}} - \frac{1}{\varepsilon_{\rm IR}} \;\;& ( n = -2 )
\end{array} \right.,
\end{equation}
where only the logarithmically divergent scaleless integral contributes the UV-divergence.

According to Lorentz covariance and oddness of the scaleless integrals, we have
\begin{equation}\label{Identity2}
\frac{\PA{2\pi\mu}^{4-D}}{i\pi^2} \int d^D k \;k^{\mu_1} k^{\mu_2} \cdots k^{\mu_m} \PA{k^2}^n = \left\{ \begin{array}{ll}
0 & ( \mbox{m is odd} ) \\
\mathcal{C}_{m,n} \times g^{\{\mu_1 \mu_2 \cdots \mu_m\}} & ( \mbox{m is even} )
\end{array} \right.,
\end{equation}
i.e. when $m$ is odd, the scaleless tensor integrals are 0, and when $m$ is even, they are proportional to $g^{\{\mu_1\mu_2\cdots\mu_m\}}$ with coefficient $\mathcal{C}_{m,n}$, where we use the same notations as Refs.~\cite{Sulyok:2006xp,Denner:2005nn} for $g^{\{\mu_1\mu_2\cdots\mu_m\}}$, which is the symmetrization of $g^{\mu_1\mu_2} g^{\mu_3\mu_4}\cdots g^{\mu_{m-1}\mu_{m}}$ with respect to the lorentz index $\mu_1,\mu_2,\cdots,\mu_m$, for example
\begin{eqnarray}
g^{\{\mu_1 \mu_2\}} &=& g^{\mu_1 \mu_2} , \nn\\
g^{\{\mu_1 \mu_2 \mu_3 \mu_4\}} &=& g^{\mu_1 \mu_2}g^{\mu_3 \mu_4} + g^{\mu_1 \mu_3}g^{\mu_2 \mu_4} + g^{\mu_1 \mu_4}g^{\mu_2 \mu_3} .
\end{eqnarray}
The coefficient $\mathcal{C}_{m,n}$ can be ready achieved by multiplying Eq.~(\ref{Identity2}) with the $\displaystyle\frac{m}{2}$ metric tensors $g_{\mu_1\mu_2} g_{\mu_3\mu_4}\cdots g_{\mu_{m-1}\mu_{m}}$ and contracting the indexes,
\begin{eqnarray}\label{Identity2-1}
\mathcal{C}_{m,n} &=& \frac{1}{D(D+2)\cdots(D+m-2)} \frac{\PA{2\pi\mu}^{4-D}}{i\pi^2} \int d^D k \PA{ k^2 }^{n+\frac{m}{2}} \nonumber\\
&=& \left\{ \begin{array}{ll}
0 & ( n+\frac{m}{2} \ne -2 ) \\
\frac{1}{D(D+2)\cdots(D+m-2)}\PB{ \frac{1}{\varepsilon_{\rm UV}} - \frac{1}{\varepsilon_{\rm IR}}} \;\;& ( n+\frac{m}{2} = -2 )
\end{array} \right.,
\end{eqnarray}
where we have used
\begin{equation}
g_{\mu_1\mu_2} g_{\mu_3\mu_4}\cdots g_{\mu_{m-1}\mu_{m}} g^{\{\mu_1 \mu_2 \cdots \mu_m\}} =  D(D+2)\cdots(D+m-2) \;,
\end{equation}
where $m$ is even, and more relations can be found in Ref.~\cite{Sulyok:2006xp}.

\section{Description of the Calculations\label{secDC}}
We define a function \texttt{Power} to get the asymptotic scaling of the polynomial of $k^2$ and $k\cdot p_i$ in the ultraviolet region, for example:
\begin{equation}
\P[k^2] = 2, \; \P[k\cdot p_i] = 1, \; \P[m^2] = 0,\; \P[k^2+k\cdot p_i - m^2] = 2
\end{equation}
and we can extend this function to rational expression of $k^2$ and $k\cdot p_i$.
\begin{equation}
\P\PB{\frac{\mathcal{N}}{\mathcal{D}}} \equiv \P[\mathcal{N}] - \P[\mathcal{D}] ,
\end{equation}
where $\mathcal{N}$ and $\mathcal{D}$ are some polynomials of $k^2$ and $k\cdot p_i$, for example
\begin{equation}
\P\PB{\frac{k\cdot p_1}{ \PA{k^2-m_0^2} \PA{(k+q_1)^2-m_1^2}^2 }} = -5 .
\end{equation}
So if \P~of the integrand in $\I$ is less than \texttt{-4}, then there will be no UV-divergent part in $\I$. i.e.
\begin{equation}
\I_{\UV} = 0, \quad ( \P[\Q]-\sum_i^N 2n_i < -4 ) .
\end{equation}
For example
\begin{equation}
\PB{ \frac{\PA{2\pi\mu}^{4-D}}{i\pi^2} \int d^D k \frac{k\cdot p_1}{ \PA{k^2-m_0^2} \PA{(k+p_1)^2-m_1^2}^2 } }_{\UV} = 0 .
\end{equation}

Now we are going to describe the calculations of UV-divergent parts of $\I$. For general integrand of $\I$, we can write
\begin{eqnarray}\label{Req}
\frac{\Q}{\N_1^{n_1} \N_2^{n_2} \cdots \N_{N}^{n_N}} &=&\frac{\N_1-\PA{\N_1-k^2}}{k^2} \frac{\Q}{\N_1^{n_1} \N_2^{n_2} \cdots \N_{N}^{n_N}} \nn\\
&=& \frac{\Q}{k^2\N_1^{n_1-1} \N_2^{n_2} \cdots \N_{N}^{n_N}} - \frac{\Q'}{k^2 \N_1^{n_1} \N_2^{n_2} \cdots \N_{N}^{n_N}}
\end{eqnarray}
with $\Q' = \Q\,\PA{\N_1 - k^2}$, and it is clear that
\begin{equation}
\P\PB{\frac{\Q'}{k^2 \N_1^{n_1} \N_2^{n_2} \cdots \N_{N}^{n_N}}} \le \P\PB{\frac{\Q}{\N_1^{n_1} \N_2^{n_2} \cdots \N_{N}^{n_N}}} -1 ,
\end{equation}
i.e. either the power of propagators: $n_i$ or the $\P$ of the integrand decreases by at least 1, if we apply this replacement once again in the last result, we have
\begin{eqnarray}\label{Req}
\frac{\Q}{\N_1^{n_1} \N_2^{n_2} \cdots \N_{N}^{n_N}} &=& \PB{ \frac{\Q}{\PA{k^2}^2\N_1^{n_1-2} \N_2^{n_2} \cdots \N_{N}^{n_N}} - \frac{\Q'}{\PA{k^2}^2 \N_1^{n_1-1} \N_2^{n_2} \cdots \N_{N}^{n_N}} } \nn\\
&&-\PB{ \frac{\Q'}{\PA{k^2}^2\N_1^{n_1-1} \N_2^{n_2} \cdots \N_{N}^{n_N}} - \frac{\Q''}{\PA{k^2}^2 \N_1^{n_1} \N_2^{n_2} \cdots \N_{N}^{n_N}} } ,
\end{eqnarray}
with $\Q'' = \Q'\,\PA{\N_1 - k^2}$, and we can see the power of propagators: $n_i$ or the $\P$ of the integrand decreases further by at least 1.

So we can apply this replacement again and again until one of the following cases happens:
\begin{itemize}
\item All the power of $\N_i$, becomes zero, i.e. only one type of propagator: $k^2$ is left.
\item The corresponding $\I_{\UV} = 0$, i.e.
$$\P\PB{\frac{\Q}{\N_1^{n_1} \N_2^{n_2} \cdots \N_{N}^{n_N}}} = \PA{\P[\Q]-\sum_i^N 2n_i} < -4 .$$
\end{itemize}
So after the recursive reduction, only one type of integration will be left:
\begin{equation}
\mathcal{I} = \frac{\PA{2\pi\mu}^{4-D}}{i\pi^2} \int d^D k \frac{\Q}{\PA{k^2}^n} \;,
\end{equation}
and it is ready to read the result according to Eqs.~(\ref{Identity1}) and (\ref{Identity2}).

We can take the following 3-point tensor integral as an example:
\begin{eqnarray}\label{example}
&&\frac{(k\cdot p_1)^3}{k^2 \PB{\PA{k+q_1}^2-m_1^2} \PB{\PA{k+q_2}^2-m_2^2}} \nn\\
&\to& \frac{(k\cdot p_1)^3}{\PA{k^2}^2 \PB{\PA{k+q_2}^2-m_2^2}} - \frac{(k\cdot p_1)^3 \PA{2k\cdot q_1+\redbox{q_1^2-m_1^2}}}{\PA{k^2}^2 \PB{\PA{k+q_1}^2-m_1^2} \PB{\PA{k+q_2}^2-m_2^2}} \nn\\
&\to& \PB{\frac{(k\cdot p_1)^3}{\PA{k^2}^3} - \frac{(k\cdot p_1)^3 \PA{2k\cdot q_2+\redbox{q_2^2-m_2^2}}}{\PA{k^2}^3 \PB{\PA{k+q_2}^2-m_2^2}}} \nn\\
&& - \PB{\frac{(k\cdot p_1)^3 \PA{2k\cdot q_1}}{\PA{k^2}^3 \PB{\PA{k+q_2}^2-m_2^2}} - \redbox{\frac{(k\cdot p_1)^3 \PA{2k\cdot q_1} \PA{2k\cdot q_1+q_1^2-m_1^2}}{\PA{k^2}^3 \PB{\PA{k+q_1}^2-m_1^2} \PB{\PA{k+q_2}^2-m_2^2}}} }\nn\\
&\to& \frac{(k\cdot p_1)^3}{\PA{k^2}^3}
 - \frac{(k\cdot p_1)^3 \PA{2k\cdot q_2}}{\PA{k^2}^4}
 - \redbox{\frac{(k\cdot p_1)^3 \PA{2k\cdot q_2} \PA{2k\cdot q_2-m_2^2}}{\PA{k^2}^3 \PB{\PA{k+q_2}^2-m_2^2}}}\nn\\
&& - \frac{(k\cdot p_1)^3 \PA{2k\cdot q_1}}{\PA{k^2}^4}
 - \redbox{\frac{(k\cdot p_1)^3 \PA{2k\cdot q_1} \PA{2k\cdot q_2-m_2^2}}{\PA{k^2}^3 \PB{\PA{k+q_2}^2-m_2^2}}} \nn\\
&\to& \frac{(k\cdot p_1)^3}{\PA{k^2}^3} - \frac{(k\cdot p_1)^3 \PA{2k\cdot q_2}}{\PA{k^2}^4} - \frac{(k\cdot p_1)^3 \PA{2k\cdot q_1}}{\PA{k^2}^4} \;,
\end{eqnarray}
where each step $\displaystyle\to$ means we take a replacement, and all the expressions framed with box have been dropped during the recursive expansion since the \P~is less than ${\tt-4}$ and will not contribute UV-divergent part.

Now it is ready to get the UV-divergent parts from the last line in Eq.~(\ref{example}), using Eqs.~(\ref{Identity1}) and (\ref{Identity2})
\begin{equation}\label{3point}
\PB{ \frac{\PA{2\pi\mu}^{4-D}}{i\pi^2}\!\!\! \int d^D k \frac{(k\cdot p_1)^3}{k^2 \PB{\PA{k+q_1}^2-m_1^2} \PB{\PA{k+q_2}^2-m_2^2}}}_{\UV} \!\!\!\!= -\frac{p_1^2 \PA{p_1\cdot q_1+p_1\cdot q_2}}{4} \frac{1}{\varepsilon_{\UV}}
\end{equation}
and we can check that it agrees with Refs.~\cite{Sulyok:2006xp,Denner:2005nn}.

The important feature of this method is that, there is only one preliminary integral, i.e. Eq.~(\ref{Identity2}), which involves no other scales like external momenta $p_i$ or mass $m_i$.

Another advantage is that it can be easily implemented in any symbolic computer language, like \textsc{Mathematica}, \textsc{Reduce}, \textsc{Form}, etc. I will give an explicit implementation with \textsc{Mathematica} in the next section.

\section{Implementation With \textsc{Mathematica}\label{secMath}}

\subsection{Typical Examples}

An implementation in \textsc{Mathematica} is already available, note that the \textsc{FeynCalc} package has been used to deal with the \texttt{ScalarProduct}, but it not required for the implementation.

The UV-divergent parts can be retrieved with the function: \verb=$UVPart=, which is defined as:
\SaveVerb{uvpart}=$UVPart[exp, k]=
\begin{equation}
\UseVerb{uvpart} := \left[ \frac{\PA{2\pi\mu}^{4-D}}{i\pi^2} \int d^D \texttt{k} \; \texttt{exp} \right]_{\UV}
\end{equation}
where \texttt{exp} can be any expression at one-loop level, while \texttt{k} is the loop momentum.

We can give some simple examples:
\begin{verbatim}
den=SPD[k] (SPD[k+q1]-m1^2) (SPD[k+q2]-m2^2)//FCI//ScalarProductExpand;
num=SPD[k, p1]^3//FCI;
$UVPart[num/den, k]//Simplify
\end{verbatim}
The output of the code above reads:
\begin{equation}\label{mout1}
\texttt{Out[ ]}:=-\frac{6 \text{p1}^2 (\text{p1}\cdot \text{q1}+\text{p1}\cdot \text{q2})}{D (D+2) \omega } \;,
\end{equation}
where $\omega$ is just $\varepsilon_{\UV}$ which we have used to represent the UV divergence in the \textsc{Mathematic} code, Eq.~(\ref{mout1}) gives the same result as Eq.~(\ref{3point}) after setting the dimension $D$ to 4.

\begin{verbatim}
den = (SPD[k]-m0^2) (SPD[k]-m1^2)^2 (SPD[k]-m2^2)//FCI//ScalarProductExpand;
num = SPD[k,p]^8//FCI;
$UVPart[num/den,k]
\end{verbatim}

The output reads:
\begin{equation}
\texttt{Out[ ]}:= \frac{105 \left(\text{m0}^4+\left(2 \text{m1}^2+\text{m2}^2\right) \text{m0}^2+3 \text{m1}^4+\text{m2}^4+2 \text{m1}^2 \text{m2}^2\right) (p^2)^4}{D \left(D^3+12 D^2+44 D+48\right) \omega } \;,
\end{equation}
setting $D$ to 4 we get
\begin{eqnarray}
&&\PB{ \frac{\PA{2\pi\mu}^{4-D}}{i\pi^2} \int d^D k \frac{(k\cdot p)^8}{\PA{k^2-m_0^2} \PA{k^2-m_1^2}^2 \PA{k^2-m_2^2}} }_{\UV} \nn\\
&=& -\frac{7 (p^2)^4 \PA{ m_0^4+m_2^4+3m_1^4+2m_0^2 m_1^2+m_0^2 m_2^2+2m_1^2m_2^2 } }{128} \frac{1}{\varepsilon_{\UV}} \;.
\end{eqnarray}

\subsection{UV-Divergences in $e^+e^- \to J/\psi + \eta_c$ at One-Loop Level\label{secExample}}
\begin{figure}[h]
\centering
\includegraphics[width=0.3\textwidth]{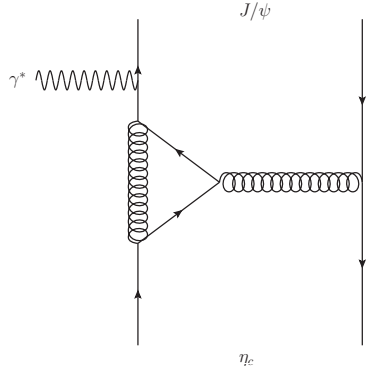}
\caption{An triangle feynman diagram in $e^+e^- \to \gamma^* \to J/\psi + \eta_c$\label{fig}}
\end{figure}

We will apply the method to a specific process $e^+e^- \to J/\psi + \eta_c$, where we use the \textsc{FeynArts} to generate Feynman diagrams, and \textsc{FeynCalc} to handle the DiracTrace. We will take a triangle Feynman diagram in Fig.~\ref{fig} as a concrete example in this section.

First we use the \textsc{FeynArts} to generate the amplitude for the diagram, then make the following replacement with the help of spinor projectors~\cite{hep-ph/0211085}:
\begin{eqnarray}
v(\bar{p}_3) \bar{u}(p_3) &\to& \frac{1}{4\sqrt{2}E_3 (E_3+m_c)} (\bar{p}\!\!\!/_3-m_c) \epsilon_S^* (P\!\!\!\!/_3+2E_3) (p\!\!\!/_3+m_c) \otimes \frac{\bf 1}{\sqrt{N_c}} \nonumber\\
v(\bar{p}_4) \bar{u}(p_4) &\to& \frac{1}{4\sqrt{2}E_2 (E_4+m_c)} (\bar{p}\!\!\!/_4-m_c) \gamma_5 (P\!\!\!\!/_4+2E_4) (p\!\!\!/_4+m_c) \otimes \frac{\bf 1}{\sqrt{N_c}}
\end{eqnarray}
where the subscripts 3 and 4 are used to label $J/\psi$ and $\eta_c$ respectively.

After performing the DiracTrace on the fermion chains, we get the amplitude for this diagram as follows:
\begin{eqnarray}\label{amp}
{\rm Amp} &=&\frac{i C_F e g_s^4 \epsilon ^{\mu \epsilon_S^* p_3 p_4} }{6 \sqrt{N_c} m_c^5 (s-4) s^2 k^2 \left(k^2+k\cdot p_4\right) \left(2 s m_c^2+k^2+k\cdot p_3+2 k\cdot p_4\right)} \nn\\
&&\times\Big(\!\!-\!\!2 D s^2 m_c^4+12 s^2 m_c^4+8 D s m_c^4-56 s m_c^4+32 m_c^4+8 D k^2 m_c^2 \nn\\
&&-2 D s k^2 m_c^2+4 s k^2 m_c^2-16 k^2 m_c^2+4 D k\cdot p_3 m_c^2+D^2 s k\cdot p_3 m_c^2 \nn\\
&&-12 D s k\cdot p_3 m_c^2 +36 s k\cdot p_3 m_c^2-16 k\cdot p_3 m_c^2+4 D k\cdot p_4 m_c^2 \nn\\
&&-D^2 s k\cdot p_4 m_c^2+10 D s k\cdot p_4 m_c^2  -32 s k\cdot p_4 m_c^2-2 D k\cdot p_4^2 \nn\\
&&+4 k\cdot p_4^2+2 D k\cdot p_3 k\cdot p_4-4 k\cdot p_3 k\cdot p_4\Big) \;,
\end{eqnarray}
where $p_3$ and $p_4$ are the momenta of $J/\psi$ and $\eta_c$ respectively, and $s=(p_3+p_4)^2/(4m_c^2)$, $m_c$ is the quark mass, and $k$ is the loop momentum, and to get the UV-divergent part of the amplitude, we just use
\begin{verbatim}
        $UVPart[Amp,k]
\end{verbatim}
The output reads:
\begin{eqnarray}
\texttt{Out[ ]}:=-\frac{i C_F (D-2)^2 e g_s^4 \epsilon ^{\gamma \psi \text{p3}\text{p4}}}{3 \sqrt{N_c} D \text{mc}^3 s^2 \omega }
\end{eqnarray}
where the Lorentz index $\gamma$ refers to $\mu$, and $\psi$ to $\epsilon_S^*$ in Eq.~(\ref{amp}). After setting $D$ to 4, we get
\begin{eqnarray}
\PB{ \frac{\PA{2\pi\mu}^{4-D}}{i\pi^2} \int d^D k \; {\rm Amp} }_{\rm UV} = -\frac{i C_F e g_s^4 \epsilon ^{\mu \epsilon^* p_3 p_4}}{3 \sqrt{N_c} m_c^3 s^2 }\frac{1}{\varepsilon_{\UV}}
\end{eqnarray}

We can apply this method to each diagram to get the corresponding UV-divergent part of the amplitude, and to check the validity of the our result. We have compared the UV-divergent part produced with our code with Ref.~\cite{Gong:2007db} diagram by diagram, and the both agree with each other for all diagrams.

\section{Summary}
A pretty simple method is introduced to calculate the UV-divergent parts at one-loop level with dimensional regularization. It is found that there is only
one preliminary integral which involves no other scale like external momenta $p_i$ or mass $m_i$ after the recursive reduction. The method can be easily implemented in any symbolic computer language, An explicit implementation with \textsc{Mathematica} is also present.

\acknowledgments
The author wants to thank Hai-Rong Dong for many useful discussions, and thanks to Xin-Qing Li for bringing me the other related fields like IR-rearrangement and also the asymptotic expansion in momenta and masses can be used for the UV extraction. The research was partially supported by China Postdoctoral Science Foundation.

\end{document}